  [2009/11/22 v0 DSPIN 2009 - proceedings write-up]
\documentclass[12pt]{article}

\usepackage{amsmath,amssymb,amsfonts,amsbsy}
\usepackage{graphicx}
\usepackage{wrapfig}
\usepackage[font=small,labelfont=bf,labelsep=period]{caption}

\usepackage{Ratcliffe}

\begin{document}

\setcounter{section}{0}
\setcounter{subsection}{0}
\setcounter{equation}{0}
\setcounter{figure}{0}
\setcounter{footnote}{0}
\setcounter{table}{0}

\begin{center}
\textbf{COLOUR MODIFICATION OF FACTORISATION\\ IN SINGLE-SPIN ASYMMETRIES}

\vspace{5mm}

\underline{Philip G. Ratcliffe}$^{\,1,\,2\,\dag}$ and Oleg V. Teryaev$^{\,3}$

\vspace{5mm}

\begin{small}
  (1) \emph{%
        Dip.to di Fisica e Matematica, Universit\`{a} degli Studi dell'Insubria, Como
      } \\
  (2) \emph{%
        Istituto Nazionale di Fisica Nucleare, Sezione di Milano--Bicocca, Milano
      } \\
  (3) \emph{%
        Bogoliubov Lab. of Theoretical Physics, Joint Inst. for Nuclear Research, Dubna
      } \\
  $\dag$ \emph{E-mail: philip.ratcliffe@unisubria.it}
\end{small}
\end{center}

\vspace{0.0mm} 

\begin{abstract}
  We discuss the way in which factorisation is partially maintained but nevertheless modified by process-dependent colour factors in hadronic single-spin asymmetries. We also examine QCD evolution of the twist-three gluonic-pole strength defining an effective T-odd Sivers function in the large-$x$ limit, where evolution of the T-even transverse-spin DIS structure function $g_2$ is known to be multiplicative.
\end{abstract}

\vspace{7.2mm}

\section{Preamble}

\subsection{Motivation}
Single-spin asymmetries (SSA's) have long been something of an enigma in high-energy hadronic physics. Prior to the first experimental studies, hadronic SSA's were predicted to be very small for a variety of reasons. Experimentally, however, they turn out to be large (up to the order of 50\% and more) in many hadronic processes. It was also long held that such asymmetries should eventually vanish with growing energy and/or $p_T^{}$. Again, however, the SSA's so far observed show no signs of high-energy suppression.

\subsection{SSA Basics}
Typically, SSA's reflect spin--momenta correlations of the form $\vec{s}\cdot(\vec{p}\vprod\vec{k})$, where $\vec{s}$ is some particle polarisation vector, while $\vec{p}$ and $\vec{k}$ are initial/final particle/jet momenta. A simple example might be: $\vec{p}$ the beam direction, $\vec{s}$ the target polarisation (transverse therefore with respect to $\vec{p}$) and $\vec{k}$ the final-state particle direction (necessarily then out of the $\vec{p}$--$\vec{s}$ plane). Polarisations involved in SSA's must usually thus be transverse (although there are certain special exceptions).

It is more convenient to use an helicity basis via the transformation
\begin{equation}
  \ket{\uparrow/\downarrow} =
  \tfrac1{\sqrtno2\,} \left[ \strut\, \ket{+} \pm \im \ket{-} \right].
\end{equation}
A transverse-spin asymmetry then takes on the (schematic) form
\begin{equation}
  \mathcal{A}_N
  \sim
  \frac{\braket{\uparrow|\uparrow}-\braket{\downarrow|\downarrow}}
       {\braket{\uparrow|\uparrow}+\braket{\downarrow|\downarrow}}
  \sim
  \frac{2\Im\braket{+|-}}{\braket{+|+}+\braket{-|-}}.
\end{equation}
The appearance of both $\ket{+}$ and $\ket{-}$ in the numerator signals the presence of a helicity-flip amplitude. The precise form of the numerator implies interference between two different helicity amplitudes: one helicity-flip and one non-flip, with a relative phase difference (the imaginary phase implying na\"{\i}ve T-odd processes).

Early on \citet{Kane:1978nd} realised that in the massless (or high-energy) limit and the Born approximation a gauge theory such as QCD cannot furnish either requirement: for a massless fermion, helicity is conserved and tree-diagram amplitudes are always real. This led to the now infamous statement \cite{Kane:1978nd}: ``\dots\ \emph{observation of significant polarizations in the above reactions would contradict either QCD or its applicability.}''

It therefore caused much surprise and interest when large asymmetries were found; QCD nevertheless survived! \citet*{Efremov:1984ip} soon discovered one way out within the context of perturbative QCD. Consideration of the three-parton correlators involved in, \emph{e.g.}\ $g_2$, leads to the following crucial observations: the relevant mass scale is not that of the current quark, but of the hadron and the pseudo-two-loop nature of the diagrams can generate an imaginary part in certain regions of partonic phase space~\cite{Efremov:1981sh}.

It took some time, however, before real progress was made and the richness of the newly available structures was fully exploited---see \cite{Qiu:1991pp.x}. Indeed, it turns out that there are a variety of mechanisms that can generate SSA's:
\begin{itemize}
\item
Transversity: this correlates hadron helicity flip to quark flip. Chirality conservation, however, requires another T-odd (distribution or fragmentation) function.
\item
Internal quark motion: the transverse polarisation of a quark may be correlated with its own transverse momentum. This corresponds to the \citeauthor{Sivers:1989cc} function \cite{Sivers:1989cc} and requires orbital angular momentum together with soft-gluon exchange.
\item
Twist-3 transverse-spin dependent three-parton correlators (\emph{cf.} $g_2$): here the pseudo two-loop nature provides effective spin flip (via the extra parton) and also the required imaginary part (via pole terms).
\end{itemize}
The second and third mechanisms turn out to be related.

\section{Single-Spin Asymmetries}

\subsection{Single-Hadron Production}
As a consequence of the multiplicity of underlying mechanisms, there are various types of distribution and fragmentation functions that can be active in generating SSA's (even competing in the same process):
\begin{itemize}
\item
higher-twist distribution and fragmentation functions,
\item
$k_T$-dependent distribution and fragmentation functions,
\item
interference fragmentation functions,
\item
higher-spin functions, \emph{e.g.}\ vector-meson fragmentation functions.
\end{itemize}
Consider then hadron production with one initial-state, transversely polarised hadron:
\begin{equation}
  A^\uparrow(P_A) + B(P_B) \to h(P_h) + X,
\end{equation}
where hadron $A$ is transversely polarised while $B$ is not. The unpolarised (or spinless) hadron $h$ is produced at large transverse momentum $\Vec{P}_{hT}$ and PQCD is thus applicable. Typically, $A$ and $B$ are protons while $h$ may be a pion or kaon \emph{etc}.
\begin{figure}[hbt]
  \centering
  \includegraphics[width=0.6\textwidth]{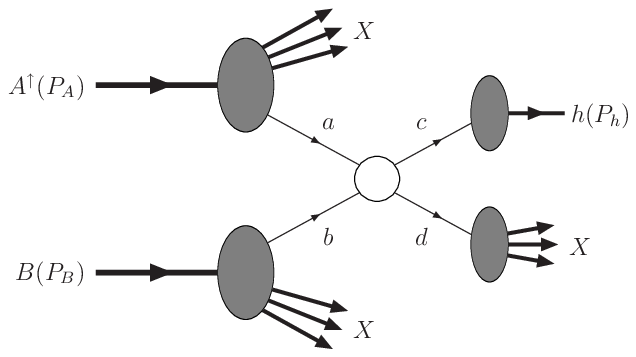}
  \caption{%
    Factorisation in single-hadron production with a transversely polarised hadron.
  }
  \label{fig:fact}
\end{figure}

\begin{figure}[hbt]
  \centering
  \raisebox{15mm}{$\mathcal{M}_{\alpha\beta\gamma\delta}$ \quad = \quad}
  \includegraphics[width=0.3\textwidth]{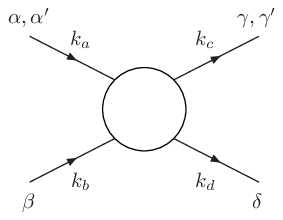}
  \caption{%
    The hard partonic amplitude, $\alpha\beta\gamma\delta$ are Dirac indices.
  }
  \label{fig:matel}
\end{figure}

The following SSA may then be measured:
\begin{equation}
  A_{T}^h =
  \frac{\D\sigma(\Vec{S}_T) - \D\sigma(-\Vec{S}_T)}
       {\D\sigma(\Vec{S}_T) + \D\sigma(-\Vec{S}_T)}.
\end{equation}
Assuming standard factorisation to hold, the differential cross-section for such a process may be written formally as (\emph{cf.} Fig.~\ref{fig:fact})
\begin{equation}
  \D\sigma =
  \sum_{abc} \sum_{\alpha\alpha'\gamma\gamma'}
  \rho^a_{\alpha'\alpha} \,
  f_a(x_a) \otimes
  f_b(x_b) \otimes
  \D\hat\sigma_{\alpha\alpha'\gamma\gamma'} \otimes
  \mathcal{D}_{h/c}^{\gamma'\gamma}(z),
\end{equation}
where $f_a$ ($f_b$) is the density of parton type $a$ ($b$) inside hadron $A$ ($B$), $\rho^a_{\alpha\alpha'}$ is the spin density matrix for parton $a$, $\mathcal{D}_{h/c}^{\gamma\gamma'}$ is the fragmentation matrix for parton $c$ into the final hadron $h$ and $\D\hat\sigma_{\alpha\alpha'\gamma\gamma'}$ is the partonic cross-section:
\begin{equation}
  \left(\frac{\D\hat\sigma}{\D\hat{t}}\right)_{\!\!\alpha\alpha'\gamma\gamma'}
  =
  \frac1{16\pi\hat{s}^2} \, \frac12 \, \sum_{\beta\delta}
  \mathcal{M}^{\vphantom*}_{\alpha\beta\gamma\delta} \,
  \mathcal{M}^*_{\alpha'\beta\gamma'\delta},
\end{equation}
where $\mathcal{M}_{\alpha\beta\gamma\delta}$ is the amplitude for the hard partonic process, see Fig.~\ref{fig:matel}.

The off-diagonal elements of $\mathcal{D}_{h/c}^{\gamma\gamma'}$ vanish for an unpolarised produced hadron; \emph{i.e.}, $\smash{\mathcal{D}_{h/c}^{\gamma\gamma'}\propto\delta_{\gamma\gamma'}}$. Helicity conservation then implies $\alpha=\alpha'$, so that there can be no dependence on the spin of hadron~$A$ and all SSA's must vanish. To avoid such a conclusion, either intrinsic quark transverse motion, or higher-twist effects must be invoked.

\subsection{Intrinsic Transverse Motion}
Quark intrinsic transverse motion can generate SSA's in three essentially different ways (all necessarily $T$-odd effects):
\begin{enumerate}
\item\label{it:Sivers}
$\Vec{k}_T$ in hadron $A$ requires $f_a(x_a)$ to be replaced by $\mathcal{P}_a(x_a,\Vec{k}_T)$, which may then depend on the spin of $A$ (distribution level);
\item\label{it:Collins}
$\Vec\kappa_T$ in hadron $h$ allows $\mathcal{D}_{h/c}^{\gamma\gamma'}$ to be non-diagonal (fragmentation level);
\item\label{it:Boer}
$\Vec{k}'_T$ in hadron $B$ requires $f_b(x_b)$ to be replaced by $\mathcal{P}_b(x_b,\Vec{k}'_T)$---the spin of $b$ in the unpolarised $B$ may then couple to the spin of $a$ (distribution level).
\end{enumerate}
The three corresponding mechanisms are: \ref{it:Sivers}.~the \citeauthor{Sivers:1989cc} effect~\cite{Sivers:1989cc}; \ref{it:Collins}.~the \citeauthor{Collins:1993kk} effect~\cite{Collins:1993kk}; \ref{it:Boer}.~an effect studied by \citeauthor{Boer:1999mm}~\cite{Boer:1999mm} in Drell-Yan processes. Note that all such intrinsic-$\Vec{k}_T$, -$\Vec\kappa_T$ or -$\Vec{k}'_T$ effects are $T$-odd; \emph{i.e.}, they require ISI or FSI. Note too that when transverse parton motion is included, the QCD factorisation theorem is not completely proven, but see~\cite{Ji:2004wu}.

Assuming factorisation to be valid, the cross-section is
\begin{multline}
  E_h \, \frac{\D^3\sigma}{\D^3\Vec{P}_h} =
  \sum_{abc} \sum_{\alpha\alpha'\beta\beta'\gamma\gamma'}
  \int \! \D{x}_a \,
  \D{x}_b \,
  \D^2\Vec{k}_T \,
  \D^2\Vec{k}'_T \,
  \frac{\D^2\Vec\kappa_T}{\pi z}
\\
  \null \times
  \mathcal{P}_a(x_a, \Vec{k}_T) \, \rho^a_{\alpha'\alpha} \,
  \mathcal{P}_b(x_b, \Vec{k}'_T) \, \rho^b_{\beta'\beta}
  \left(
    \frac{\D\hat\sigma}{\D\hat{t}}
  \right)_{\alpha\alpha'\beta\beta'\gamma\gamma'}
  \mathcal{D}_{h/c}^{\gamma'\gamma}(z, \Vec\kappa_T),
\end{multline}
where again
\begin{equation}
  \left(
    \frac{\D\hat\sigma}{\D\hat{t}}
  \right)_{\alpha\alpha'\beta\beta'\gamma\gamma'}
  =
  \frac1{16\pi\hat{s}^2} \, \sum_{\beta\delta}
  \mathcal{M}_{\alpha\beta\gamma\delta} \,
  \mathcal{M}^*_{\alpha'\beta\gamma'\delta}.
\end{equation}
The Sivers effect relies on $T$-odd $k_T$-dependent distribution functions and predicts an SSA of the form
\begin{multline}
  E_h \, \frac{\D^3\sigma( \Vec{S}_T)}{\D^3\Vec{P}_h} -
  E_h \, \frac{\D^3\sigma(-\Vec{S}_T)}{\D^3\Vec{P}_h}
\\
  \null =
  | \Vec{S}_T | \,
  \sum_{abc} \int \! \D{x}_a \, \D{x}_b \, 
  \frac{\D^2\Vec{k}_T}{\pi z} \Delta_0^T{f}_a(x_a, \Vec{k}_T^2) \, f_b(x_b) \, 
  \frac{\D\hat\sigma(x_a, x_b, \Vec{k}_T)}{\D\hat{t}} \, D_{h/c}(z),
\end{multline}
where $\Delta_0^T{f}$ (related to $f_{1T}^\perp$) is a $T$-odd distribution.

\subsection{Higher Twist}
\citet*{Efremov:1984ip} showed that in QCD non-vanishing SSA's can also be obtained by invoking higher twist and the so-called gluonic poles in diagrams involving $qqg$ correlators. Such asymmetries were later evaluated in the context of QCD factorisation by \citeauthor*{Qiu:1991pp.x}, who studied both direct-photon production \cite{Qiu:1991pp.x} and hadron production~\cite{Qiu:1998ia}. This program has been extended by \citet*{Kanazawa:2000hz.x} to the chirally-odd contributions. The various possibilities are:
\begin{align}
  \D\sigma &=
  \sum_{abc}
  \left\{ \vphantom{D_{h/c}^{(3)}}
    G_F^a(x_a, y_a) \otimes f_b(x_b) \otimes \D\hat\sigma \otimes D_{h/c}(z)
  \right.
  \nonumber
\\
  & \hspace{8em} \null +
  \DT{f}_a(x_a) \otimes E_F^b(x_b, y_b) \otimes \D\hat\sigma' \otimes D_{h/c}(z)
  \nonumber
\\[1ex]
  & \hspace{14em} \null +
  \left.
    \DT{f}_a(x_a) \otimes f_b(x_b) \otimes \D\hat\sigma'' \otimes D_{h/c}^{(3)}(z)
  \right\}.
\end{align}
The first term is the chirally-even three-parton correlator pole mechanism, as proposed in \cite{Efremov:1984ip} and studied in \cite{Qiu:1991pp.x,Qiu:1998ia}; the second contains transversity and is the chirally-odd contribution analysed in \cite{Kanazawa:2000hz.x}; and the third also contains transversity but requires a \mbox{twist-3} fragmentation function $D_{h/c}^{(3)}$.

\subsection{Phenomenology}
\citet{Anselmino:2002pd} have compared data with various models inspired by the previous possible ($k_T$-dependent) mechanisms and find good descriptions although they were not able to differentiate between contributions. The calculations by \citet*{Qiu:1991pp.x} (based on three-parton correlators) also compare well but are rather complex. However, the \mbox{twist-3} correlators (as in $g_2$) obey constraining relations with $k_T$-dependent densities and also exhibit a novel factorisation property, to which we now turn.

\subsection{Pole Factorisation}
\begin{figure}[b]
  \centering
  \includegraphics[height=25mm,bb=142 661 250 749,clip]{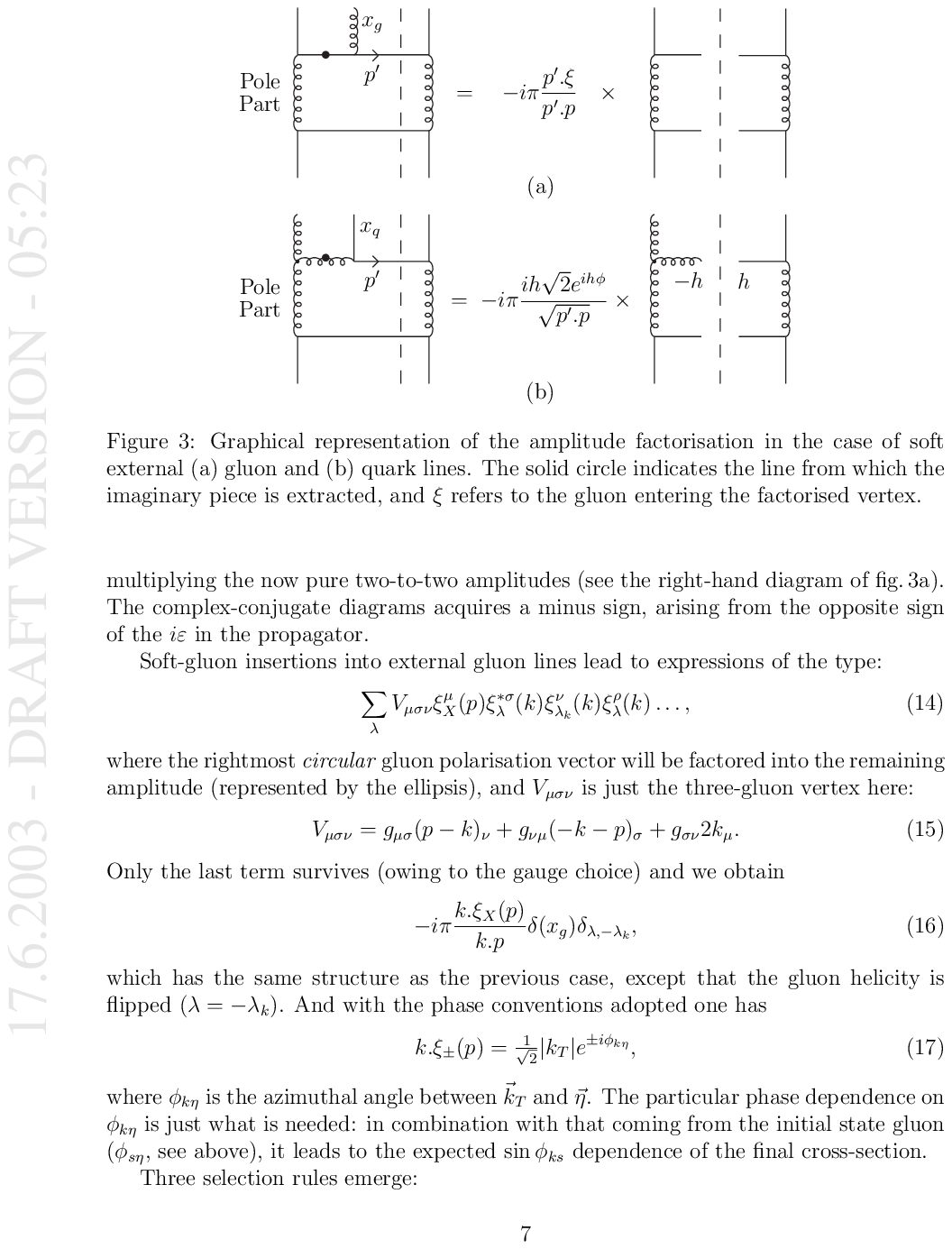}
  \includegraphics[height=25mm,bb=250 661 437 749,clip]{epsfiles/pole-factor}
  \caption{%
    An example of pole factorisation: $p$ is the incoming proton momentum, $p'$ the outgoing hadron and $\xi$ is the gluon polarisation vector (lying in the transverse plane).
  }
  \label{fig:polefact}
\end{figure}

\citet*{Efremov:1984ip} noticed that the \mbox{twist-3} diagrams involving three-parton correlators can supply the necessary imaginary part via a pole term; spin-flip is implicit (and
\begin{wrapfigure}[12]{l}{40mm}
  \vspace*{-4mm}
  \centering
  \includegraphics[width=30mm,bb=92 558 196 701,clip]{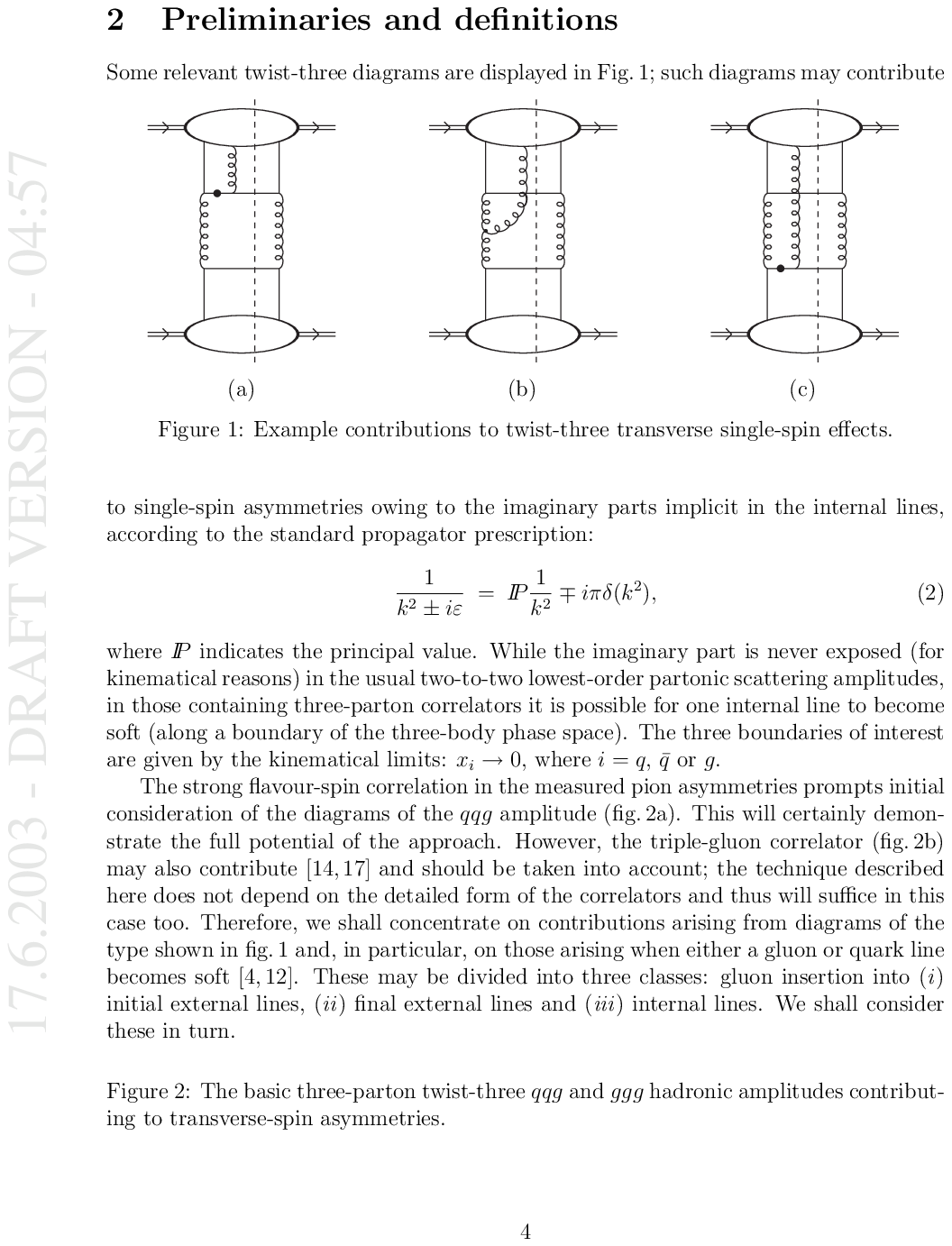}
  \caption{Example of a propagator pole in a three-parton diagram.}
  \label{fig:propagator-pole}
\end{wrapfigure}
due to the gluon). The standard propagator prescription ($-\mkern-7mu\bullet\mkern-7mu-$ in Fig.~\ref{fig:propagator-pole}, with momentum $k$),
\begin{equation}
  \frac{1}{k^2\pm\im\varepsilon} = \Principal \frac1{k^2} \mp \im\pi\delta(k^2).
\end{equation}
leads to an imaginary contribution for $k^2\to0$. A gluon with momentum $x_gp$ inserted into an (initial or final) external line $p'$ sets $k=p'-x_gp$ and thus as $x_g\to0$ we have that $k^2\to0$. The gluon vertex may then be factored out together with the quark propagator and pole, see Fig.~\ref{fig:polefact}.
Such factorisation can be performed systematically for all poles (gluon and fermion): on all external legs with all insertions \citep{Ratcliffe:1998pq}. The structures are still complex: for a given correlator there are many insertions, leading to different signs and momentum dependence.

\begin{wrapfigure}[13]{r}{40mm}
  \centering
  \includegraphics[width=30mm,bb=391 558 495 701,clip]{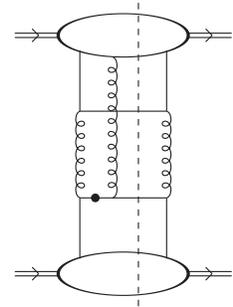}
  \caption{Example of a dominant propagator pole diagram.}
  \label{fig:dominant}
\end{wrapfigure}

The colour structures of the various diagrams (with the different types of soft insertions) are also different (we shall examine this question shortly). In all cases (examined) it turns out that just one diagram dominates in the large-$\Nc$ limit, see Fig.~\ref{fig:dominant}. All other insertions into external (on-shell) legs are relatively suppressed by $1/\Nc^2$. This 
has been examined in detail by \citeauthor{Ramilli:2007t1} (Insubria U. Masters thesis \cite{Ramilli:2007t1}): the leading diagrams provide a good approximation. The analysis has yet to be repeated for all the other \mbox{twist-3} contributions (\emph{e.g.}\ also in fragmentation).

A question immediately arises: could there be any direct relationship between the \mbox{twist-3} and $k_T$-dependent mechanisms? It might be hoped that, via the equations of motion \emph{etc.}, unique predictions for single-spin azimuthal asymmetries could be obtained by linking the (\emph{e.g.}\ Sivers- or Collins-like) $k_T$-dependent mechanisms to the (Efremov--Teryaev) higher-twist three-parton mechanisms. An early attempt was made by \citet*{Ma:2003ut} for the Drell-Yan process, but the predictions were found not to be unique. \citet{Ji:2006ub.x} have since also examined the relationship between the $k_T$-dependent and higher-twist mechanisms by matching the two in an intermediate $k_T$ region of common validity.

\section{More on Multiparton Correlators}

\subsection{Colour Modification}
In \cite{Ratcliffe:2007ye} we provided an \emph{a posteriori} proof of the relation between \mbox{twist-3} and $k_T$-dependence. The starting point is a factorised formula for the Sivers function:
\begin{align}
  d\Delta\sigma
  & \sim
  \int \! d^2k_T^{} dx \, 
  f_{\text{S}}^{} (x,k_T^{}) \, \epsilon^{\rho s P k_T^{}}
  \Tr\!\left[\vphantom{\big|} \gamma_\rho \, H(xP,k_T^{}) \right].
\\
\intertext{%
  Expanding the subprocess coefficient function $H$ in powers of $k_T^{}$ and keeping the first non-vanishing term leads to
}
  & \sim
  \int \!\! d^2k_T^{} dx \,
  f_{\text{S}}^{} (x,k_T^{}) \, k_T^\alpha \, \epsilon^{\rho s P k_T^{}}
  \Tr\!\left[
    \gamma_\rho \, \frac{\partial H(xP,k_T^{})}{\partial k_T^\alpha}
  \right]_{k_T^{}{=}0}
  \mkern-35mu.
\end{align}
By exploiting various identities and the fact that there are other momenta involved, this can be rearranged into the following form:
\begin{equation}
  d\Delta\sigma \sim
  M \int \! dx \, f_{\text{S}}^{(1)}(x) \, \epsilon^{\alpha s P n}
  \Tr\!\left[
    \slashed{P} \, \frac{\partial H(xP,k_T^{})}{\partial k_T^\alpha}
  \right]_{k_T^{}{=}0}
  \mkern-35mu,
\end{equation}
where
\begin{equation}
  f_{\text{S}}^{(1)}(x)
  = \int \! d^2k_T^{} \, f_{\text{S}}^{}(x,k_T^{}) \, \frac{k_T^2}{2M^2}.
\end{equation}
The final expression coincides with the master formula of \citet*{Koike:2006qv.x} for \mbox{twist-3} gluonic poles in high-$p_T^{}$ processes. The Sivers function can thus be identified with the gluonic-pole strength $T(x,x)$ multiplied by a process-dependent colour factor.

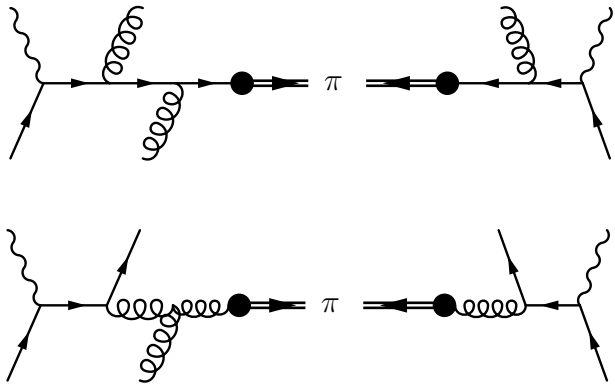
\begin{figure}[hbt]
  \centering
  \begin{fmffile}{fmffiles/SiversCol}%
    \fmfset{arrow_len}{2.5mm}%
    \fmfset{curly_len}{2.2mm}%
    \begin{fmfgraph*}(44,20)
      \fmfleft{i1,i2}
      \fmfright{d1,d2,d3}
      \fmftop{o2}
      \fmfbottom{o1}
      \fmf{phantom}{i1,v1,v2,o1}
      \fmf{phantom}{i2,v1}
      \fmf{phantom}{v2,o2}
      \fmffreeze
      \fmf{fermion}{i1,v1,v2,v3,v4}
      \fmf{dbl_plain_arrow}{v4,d2}
      \fmffreeze
      \fmf{photon}{v1,i2}
      \fmf{gluon}{o2,v2}
      \fmf{gluon}{o1,v3}
      \fmfv{d.sh=circle,d.fi=full,d.si=4thick}{v4}
      \fmfv{l=$\pi$}{d2}
    \end{fmfgraph*}
    \hspace{1.2em}
    \begin{fmfgraph}(36,20)
      \fmfright{i1,i2}
      \fmfleft{d1,d2,d3}
      \fmftop{o2}
      \fmfbottom{o1}
      \fmf{phantom}{i1,v1,v2,o1}
      \fmf{phantom}{i2,v1}
      \fmf{phantom}{v2,o2}
      \fmffreeze
      \fmf{fermion}{i1,v1,v2,v4}
      \fmf{dbl_plain_arrow}{v4,d2}
      \fmffreeze
      \fmf{photon}{v1,i2}
      \fmf{gluon}{v2,o2}
      \fmfv{d.sh=circle,d.fi=full,d.si=4thick}{v4}
    \end{fmfgraph}
    \\[5ex]
    \hspace*{0.15em}
    \begin{fmfgraph*}(44,20)
      \fmfleft{i1,i2}
      \fmfright{d1,d2,d3}
      \fmftop{o2}
      \fmfbottom{o1}
      \fmf{phantom}{i1,v1,v2,o1}
      \fmf{phantom}{i2,v1}
      \fmf{phantom}{v2,o2}
      \fmffreeze
      \fmf{fermion}{i1,v1,v2}
      \fmf{gluon}{v2,v3,v4}
      \fmf{dbl_plain_arrow}{v4,d2}
      \fmffreeze
      \fmf{photon}{v1,i2}
      \fmf{fermion}{v2,o2}
      \fmf{gluon}{o1,v3}
      \fmfv{d.sh=circle,d.fi=full,d.si=4thick}{v4}
      \fmfv{l=$\pi$}{d2}
    \end{fmfgraph*}
    \hspace{1.2em}
    \begin{fmfgraph}(36,20)
      \fmfright{i1,i2}
      \fmfleft{d1,d2,d3}
      \fmftop{o2}
      \fmfbottom{o1}
      \fmf{phantom}{i1,v1,v2,o1}
      \fmf{phantom}{i2,v1}
      \fmf{phantom}{v2,o2}
      \fmffreeze
      \fmf{fermion}{i1,v1,v2}
      \fmf{gluon}{v4,v2}
      \fmf{dbl_plain_arrow}{v4,d2}
      \fmffreeze
      \fmf{photon}{v1,i2}
      \fmf{fermion}{v2,o2}
      \fmfv{d.sh=circle,d.fi=full,d.si=4thick}{v4}
    \end{fmfgraph}
  \end{fmffile}
  \caption{Twist-3 SIDIS $\pi$ production via quark and gluon fragmentation.}
  \label{fig:SIDIS-3}
\end{figure}

The sign of the Sivers function depends on which of ISI or FSI is relevant:
\begin{equation}
  f_{\text{S}}^{(1)}(x) = \sum_i C_i \, \frac{1}{2M} T(x,x),
\end{equation}
where $C_i$ is a relative colour factor defined with respect to an Abelian subprocess. Emission of an extra hard gluon is needed to generate high $p_T^{}$ and, according to the process under consideration, FSI may occur \emph{before} or \emph{after} this emission, leading to different colour factors. In this sense, factorisation is broken in SIDIS, albeit in a simple and accountable manner. Figure~\ref{fig:SIDIS-3} depicts the application of this relation to high-$p_T^{}$ SIDIS.

\subsection{Asymptotic Behaviour}
The relation between gluonic poles (\emph{e.g.}\ the Sivers function, and T-even transverse-spin effects, \emph{e.g.}\ $g_2$ \citep{Shuryak:1981pi, Efremov:1983eb, Bukhvostov:1984as, Ratcliffe:1985mp, Balitsky:1987bk}) still remains unclear. Although there are model-based estimates and approximate sum rules, the compatibility of general \mbox{twist-3} evolution \nocite{Bukhvostov:1984as, Ratcliffe:1985mp, Balitsky:1987bk} with dedicated studies of gluonic-pole evolution (\cite{Kang:2008ey,Zhou:2008mz} and at NLO \cite{Vogelsang:2009pj}) is still unproven.

In the large-$x$ limit the evolution equations for $g_2$ diagonalise in the double-moment arguments \citep{Ali:1991em}. For the Sivers function and gluonic poles, this is the important kinematical region~\citep{Qiu:1991pp.x}. The gluonic-pole strength $T(x)$, corresponds to a specific matrix element \citep{Qiu:1991pp.x}. It is also the residue of a general $qqg$ vector correlator $b_V(x_1,x_2)$~\cite{Korotkiian:1995vf}:
\begin{equation}
  b_V(x_1,x_2) = \frac{T(\frac{x_1+x_2}{2})}{x_1-x_2} + \text{regular part},
\end{equation}
which is defined as
\begin{equation}
  b_V(x_1,x_2) =
  \frac{i}{M} \! \int \! \frac{d\lambda_1 d\lambda_2}{2\pi} \,
  e^{i\lambda_1(x_1-x_2)+i\lambda_2 x_2} \, \epsilon^{\mu s p_1 n} \!
  \braket{p_1,s|
    \bar\psi(0) \, \slashed{n} \, D_\mu(\lambda_1) \, \psi(\lambda_2)
  |p_1,s}.
\end{equation}
There is though one other correlator, projected onto an axial Dirac matrix:
\begin{equation}
  b_A(x_1,x_2) =
  \frac1{M} \! \int \! \frac{d\lambda_1 d\lambda_2}{\pi}
  e^{i\lambda_1(x_1-x_2)+i\lambda_2x_2}
  \braket{p_1,s|
    \bar\psi(0) \, \slashed{n} \, \gamma^5 \, s{\cdot}D(\lambda_1) \, \psi(\lambda_2)
  |p_1,s}.
\end{equation}
This last is required to complete the description of transverse-spin effects, in both SSA's and $g_2$. The two correlators have opposite symmetry properties for $x_1\leftrightarrow{}x_2$:
\begin{equation}
  b_A(x_1,x_2) =  b_A(x_2,x_1), \qquad
  b_V(x_1,x_2) = -b_V(x_2,x_1),
\end{equation}
determined by $T$ invariance. In both DIS and SSA's just one combination appears~\citep{Efremov:1983eb}:
\begin{equation}
  b_-(x_1,x_2) = b_A(x_2,x_1) - b_V(x_1,x_2).
\end{equation}
The QCD evolution equations \citep{Bukhvostov:1984as, Ratcliffe:1985mp, Balitsky:1987bk} are best expressed in terms of another quantity, which is determined by matrix elements of the gluon field strength:
\begin{equation}
  Y(x_1,x_2) = (x_1-x_2) \, b_-(x_1,x_2).
\end{equation}
It should be safe to assume that $b_-(x_1,x_2)$ has no double pole and
thus
\begin{equation}
  T(x) = Y(x,x).
\end{equation}

Evolution is easier studied in Mellin-moment form, implying double moments for $Y(x,y)$:
\begin{equation}
  Y^{mn} = \int dx \, dy \, x^m \, y^n \, Y(x,y),
\end{equation}
where the allowed regions are $|x|$, $|y|$ and $|x-y|<1$ (recall that negative values indicate antiquark distributions). We wish to examine the behaviour for $x$ and $y$ both close to unity and therefore close to each other. The gluonic pole thus provides the dominant contribution:
\begin{equation}
  \lim_{x,y\to1} Y(x,y) = T(\tfrac{x+y}{2}) + O(x-y).
\end{equation}
In this approximation (now large $m,n$) the leading-order evolution equations simplify:
\begin{equation}
  \frac{d}{ds} Y^{nn} = 4\left(\CF+\frac{\CA}{2}\right) \ln{n} \; Y^{nn},
\end{equation}
where the evolution variable is $s=\beta_0^{-1}\ln\ln{Q^2}$.
In terms of $T(x)$ this is
\begin{equation}
  \dot T(x) =
  4\left(\CF+\frac{\CA}{2}\right)
  \int_x^1 dz \, \frac{(1-z)}{(1-x)} \frac1{\,(z-x)_+\!} \, T(z),
\end{equation}
which is similar to the unpolarised case, but differs by a colour factor $\left(\CF+\sfrac{\CA}{2}\right)$ and a softening factor $(1-z)/(1-x)$.

The extra piece in the colour factor ($\CA/2$) \emph{vis-a-vis} the unpolarised case (just $\CF$) reflects the presence of a third active parton---the gluon. That is, the pole structure of three-parton kernels is identical, but the effective colour charge of the extra gluon is $\CA/2$. The softening factor is inessential to the asymptotic solution, it merely implies standard evolution for the function $f(x)=(1-x)T(x)$. For an initial $f(x,Q_0^2)=(1-x)^a$, the asymptotic solution \citep{Gross:1974fm} is the same but modified by $a\to{}a(s)$, with
\begin{equation}
  a(s) = a + 4\left(\CF+\frac{\CA}{2}\right) s.
\end{equation}
For $T(x)$, $a$ shifts to $a-1$; the evolution modification is identical; the spin-averaged asymptotic solutions are thus also valid for $T(x)$. This large-$x$ limit of the evolution agrees, barring the colour factor itself, with studies of gluonic-pole evolution~\citep{Kang:2008ey, Zhou:2008mz, Vogelsang:2009pj}. We should note here that \citet{Braun:2009mi} have found errors in \cite{Kang:2008ey} and \cite{Vogelsang:2009pj}; our results for the logarithmic term are, however, unaffected.

\section{Summary and Conclusions}

Viewing the Sivers function as an effective \mbox{twist-3} gluonic-pole contribution \cite{Ratcliffe:2007ye}, it is seen to be process dependent: besides a sign (ISI \emph{vs.}\ FSI), there is a process-dependent colour factor. This factor is determined by the colour charge of the initial and final partons. It generates the sign difference between SIDIS and Drell-Yan at low $p_T^{}$, but in hadronic reactions at high $p_T^{}$ it is more complicated. Such a picture is complementary to the matching in the region of common validity. The matching between various $p_T^{}$ regions now takes the form of a $p_T^{}$-dependent colour factor. It also lends some justification to the possibility of global Sivers function fits~\citep{Teryaev:2005bp}.

We have shown that generic \mbox{twist-3} evolution is applicable to the Sivers function. Its effective nature allows us to relate the evolution of T-odd (Sivers function) and T-even (gluonic pole) quantities. A vital ingredient here is the large-$x$ approximation, where gluonic-poles dominate and the evolution simplifies. The Sivers-function evolution is then multiplicative and described by a colour-factor modified \mbox{twist-2} spin-averaged kernel~\cite{Ratcliffe:2009r1}.

\section*{Acknowledgments}

I wish to thank the organisers for inviting me to this delightful and stimulating meeting: despite their repeated invitations, this is the first time I have been able to participate in a DSPIN workshop and to visit Dubna. The studies presented here have been performed in collaboration with Oleg Teryaev, see \cite{Ratcliffe:2007ye} and work in progress \cite{Ratcliffe:2009r1}. The support for the visits of Teryaev to Como was provided by the Landau Network (Como) and also by the recently completed (and hopefully to be renewed) Italian ministry-funded PRIN2006 on Transversity. Some of the ideas have already been presented at other workshops \cite{Ratcliffe:2007p1, Ratcliffe:2008p2, Ratcliffe:2009pp}.


\end{document}